# Malicious node aware wireless multi hop networks: a systematic review of the literature and recommendations for future research

Shahram pourdehghan[1], Nahideh derakhshanfard[1*]

**Abstract:** wireless communication provides great advantages that are not available through their wired counterparts such as flexibility, ease of deployment and use, cost reductions, and convenience. Wireless multi-hop networks (WMN) do not have any centralized management infrastructure. Wireless multi-hop networks have many benefits since proposed. In such networks when a node wants to send a packet to a destination where is not in the transmission range, depend on some intermediate nodes. In this type of networks packet sending is in the form of multiple hop until destination and this work is dynamic. Lack of centralized management cause that some nodes show malicious function. Malicious nodes are that receive packets and drop them maliciously. These malicious nodes could have many reasons such as hardware failure, software failure or lack of power. Such nodes make multiple packets drop from the network and the performance of network strongly decreases. As a result, the throughput of the network decrease, increase end-to-end delay and increase overhead. Therefore, we must aware from presence of malicious node in the network and do routing based on this awareness. Therefore, this paper aims to study and review the present malicious node detection methods that proposed in literatures. We categorized networks in groups, including ad hoc networks, MANET, DTN, Opportunistic networks, WSN, VANET and other wireless networks and compare malicious node detection methods in these networks. In addition, detailed comparison of these methods in each group can help for future studies.

*Corresponding author: Nahideh Derakhshanfard
n.derakhshanfard@iaut.ac.ir

[1] Department of Computer Engineering, Tabriz Branch,
Islamic Azad University, Tabriz, Iran

# 1. Introduction

In WMN, the nodes of the network used as relay node. In such networks every node where wants to send a packet to out of itself range, use relay nodes. The independency of nodes depending on intermediate nodes. This network in known as WMNs and have various types like ad hoc networks, MANET, VANET, DTN, WSN, Opportunistic network where we explain these networks in the following.

**Hybrid P2P Networks:** in peer-to-peer networks, interconnected nodes share resources amongst each other without the use of a centralized administrative system. In a hybrid P2P network, all the nodes classified into two categories, one is the super node and another is the ordinary node (1). Under each super node, there are several ordinary nodes with which the super node forms a subnet (1).

**Mobile Ad hoc Networks (MANET):** it is a complex system where the nodes are distributed and dynamically in characteristic. They do not depend on any centralized administrator for executing their function. They do not depend on any central authority for taking any action (2). They are self-healing and self-configurable because the topology of MANET changes periodically and nodes in the MANET is free to go away anywhere in the network (2).

**Cognitive Networks:** cognitive networks are self-observant and self-healing by nature. Cognitive networks can be established to address numerous problems that exist like provisioning of Quality of Service (QoS), access control, resource management, security and other desired network goals (3).

**Heterogeneous and homogeneous wireless networks:** a heterogeneous network is network connecting computers and other devices with different operating systems and/or protocols. In these networks, nodes are not the same and they are different in term of power consumption, size, protocol, and so on.

**Ad hoc networks:** ad hoc networks are multi-hop network in which there is no need of central administration. Each node in ad hoc networks act as router to send and receive data (4).

**Ad hoc grid environment:** in ad hoc grid environment, resources are not always available since nodes can spontaneously connect and disconnect at any time. Thus, this environment demands the correct execution of tasks to guarantee good performance (5).

**Wireless sensor networks (WSN):** a typical wireless sensor network consists of a large number of small, inexpensive, resource constrained sensor nodes that communicate wirelessly in a multi-hop network. These networks are formed by easy and cheap deployment features with limited resources (6). The main function of WSN is gathering data from environments and forwarding them to the base station (6).

**Cluster based sensor networks:** cluster based hierarchical sensor networks have cluster heads (CHs), which gather, aggregate, and relay sensed data from localized member nodes (MNs) to a base station (BS) (7). Sensor nodes constitute a hierarchical architecture and work with self-organization management mechanism to reduce the energy overhead (7).

**Mobile wireless sensor networks (MWSN):** mobile wireless sensor networks (MWSN) are composed of a large number of wireless sensors and they require a careful consumption of available energy to prolong the life of the network (8).

**Vehicular Ad hoc Network (VANET):** a vehicular ad hoc network is the newest paradigm of wireless multi-hop network. It emerged from MANETs with the mobile nodes being the vehicles on the roads. Vehicular ad hoc network is a self-organized network, which connects vehicles on the road with each other and with roadside units (RSU) to improve and maintain the safety on road and to manage traffic (9).

**Wireless mesh networks:** wireless mesh network structure is similar to MANET. Wireless mesh network and MANET have to apply policy of high-level security because of the ad hoc nature of these networks (10). Wireless mesh network structure presented in figure 1.

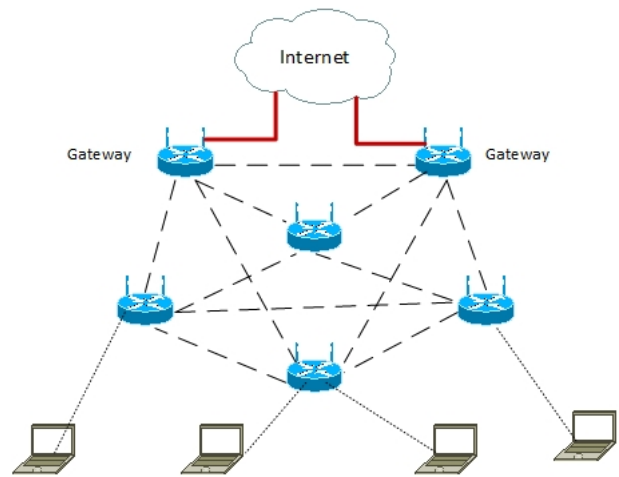

Fig. 1: wireless mesh network

**DTN:** in delay tolerant networks, the information transferred from its source to the destination without end-to-end connectivity of the network. When transmitting data between source and destination in networking, DTNs have very high transmission delay. Unlike MANETs, DTNs characterized by intermittent connectivity between nodes (11).

**Body area networks (BAN):** in recent years, wireless body area networks emerge as a key technology to support various telehealth applications. A BAN is a wireless network that is generally composed of small wearable implantable sensor nodes that placed in, on or around a patient's body (12).

**Opportunistic networks:** opportunistic networks emerged from DTN; connectivity in Opp-Nets between mobile wireless nodes is intermittent. The exchange and forward of data through opportunistic contacts between nodes, so a direct path to the destination not known. Opp-Nets differ from DTNs in that the message is always sent opportunistically, and an existing end-to-end path is never checked (13).

**Malicious node definition:** In literatures, various definition introduced about malicious node. For example, (14) noted if a neighboring node has a trust level lower than a predefined Threshold value, it identified as malicious and it not considered for route selection. (15) Said that some nodes in order to save some resources stop cooperating in the routing function. We concluded that such nodes receive packets from neighbors and drop them or uncooperative in routing operation.

**Systematic review:** a systematical literature review (SLR) is a complete study to analysis and interpretation all works that done about related research topic. This review can help researcher to know what done related a topic and help them in future works. Malicious nodes have significant effect in network performance and there is not any complete review about it. Therefore, we decided to study malicious node detection methods in previous years and help future works.

Our main works in the following:

- A systematic review about methods that proposed for malicious node detection.
- Main challenges in detecting malicious nodes in WMNs.
- Comparing performance of proposed works.

The remaining of this paper organized as follows. Section 2 explains main concepts of malicious node and WMNs. The related reviews and survey papers discussed in section 3. In section 4, research methodology explained. Section 5 describes malicious node detection methods in groups. In section 6, the conclusion of this paper presented and we summarize future works and open research issues.

## 2. Main concepts of malicious nodes and WMNs

In this section, we discussed the main concepts of wireless multi-hop networks and main concepts of malicious nodes.

### 2.1. Wireless multi-hop networks

Wireless nodes, in order to communicate with out of range nodes when wireless nodes deployed in an ad hoc setup with no infrastructure, a wireless node has to depend on other intermediate nodes for relaying its messages until they reach the intended destination. This communication paradigm known as "multi-hop" communication, where each node can act as a source, a destination, or a router relaying messages. Nodes in a wireless multi-hop network collaboratively determine their transmission power and define the network topology by forming the proper neighbor relation under the constraint of the network connectivity and the criteria with respect to energy efficiency (16).

### 2.2. Malicious node concept

Malicious nodes have different definitions in the literatures. Malicious node is at times entitled as a black hole node (2). For example, (2) noted that malicious node is a node in the network, which does flooding and leads to denial of service attack. (17) said that there are some nodes in the network whose objective is to cause harm and bring disorder to the network, these nodes referred as malicious node. These nodes do not reveal their identities while disrupting network services (17). The goal of such nodes is maximize the damage in the network. Malicious nodes may change the result of tasks and thus affect network performance (5). In some cases, an attacker can create a malicious node and spy from the network. We conclude that a malicious node in network is an uncooperative node that does not help in forwarding packet to destination. The reason of this behavior can be hardware failure, software failure, broken, battery life and/or intrusion.

### 2.3. Criterions and parameters

- *FPR*: There are many criterions in the literatures that described by author. False negative rate and false positive rate is a criterion that described in (1). False positive rate (FPR) in the ration of nodes that are normal but considered as malicious to all the normal nodes. In addition, false negative rate (FNR) is the ratio of nodes that are malicious but considered as normal to all the malicious nodes (1). Another criterion is detection error that used in (3). The article aims are reducing the malicious node detection error and presented in percentage.
- *Throughput*: Another criterion that used widely in the literature is throughput. Throughput characterized in bits per second (bps) and it denotes the number of packets that received from the intended sender node in per unit of time (2). It is obvious that the elder throughput is required. This criterion correlated with packet delivery ratio (PDR) and end-to-end delay. Packet delivery ratio is the fraction regarding the data packets expected by the target node for those sent by equivalent source node (2). End to end delay is the measure of the time, which employed to sending packets from source to their corresponding destination to get those packets (2).
- *Attack gain*: is another criterion that used by (17). A large attack gain means more payoff gained from an attack (17). Channel unreliability is a criterion that decreases with decrease of attack gain. With large detection gain, the payoffs for node will not increase. Another criterion that used by this article is attack success rate that wants to low this criterion.

One of obvious criterion in articles is identifying and excluding malicious nodes. (18) Uses these criterions and wants to identify and exclude a misbehaving node from the network.

Some articles like (7) want to control and reduce storage, computation and communication overhead. (19) Focused exactly on the percentage of detected nodes as malicious. Detection accuracy is another criterion that articles attempt to increase it. The success rate represents the percentage of the detection scheme that can correctly identify the malicious node (8).

## 3. Related Work

There are not many survey articles about malicious node in the literatures. We found some articles that are not reliable completely and are not systematic. These articles only reviewed some proposed methods in malicious node detection or removal. For example, (20) describes malicious nodes only from one aspect and only in MANET and do not attention to other types of network. (21) Classified previous works about malicious node based on methods like trust evaluation, data mining, ant colony, agent, neighbor based that are not complete and has not used other type of classifications. (22) Has a comparative review of techniques used in malicious node detection. In addition, this article as others do not used other type of classifications. Based on our study some of disadvantages of these articles are listed as below:

- Addressing to one type of network.
- Not having complete classification based of different topics.
- Lack of complete comparison based on different methods.
- Article selection process not mentioned.
- The articles not compared.
- Previous works in this era are so brief.

## 4. Research methodology

In this section the study methodology and the question that caused to do this research is present. First, we explain question formalization and then describe how articles selected. At the end of this section, we classify articles based on publishers and year that presented about this area.

### 4.1. Question Formalization

This section addresses to identify the impact of malicious node presence in the WMNs.

Q1: what is the concept of malicious node?

This question wants to answer the main concept of malicious node in WMNs and describe a node what must do to called malicious node.

Q2: what methods presented in the literature about malicious node detection?

This question wants to present methods that proposed for malicious node detection in recent years in the literature. The answers of this question can present comparative ability about proposed methods and new ideas about future methods.

Q3: what is the impact of malicious node to the criterions like throughput, energy consumption and other parameters that are important in WMNs?

With the answer of this question, we could understand the important parameters in WMNs. In addition, with controlling these parameters in WMNs, we could improve the performance of the network.

Q4: how to select papers to gain best result of above mentioned questions?

The answer of this question will be discussed in future section.

### 4.2.     Article Selection Process

First step in article selection is automatic search using search engines like google scholar with keywords. The keywords that used for this purpose include malicious node, malicious node detection, malicious nodes in WMNs, misbehavior nodes remove and so on.   The article that we found mainly published from three publishers Elsevier, Springer and IEEE. The articles are from journals and conferences. The distribution of the obtained articles by year is shown in figure2. Figure3 shows articles based on publishers. In article selection, we attend the quality of publisher and presence of keyword in title and abstract. Figure4 compare articles based on journal article and conference article.

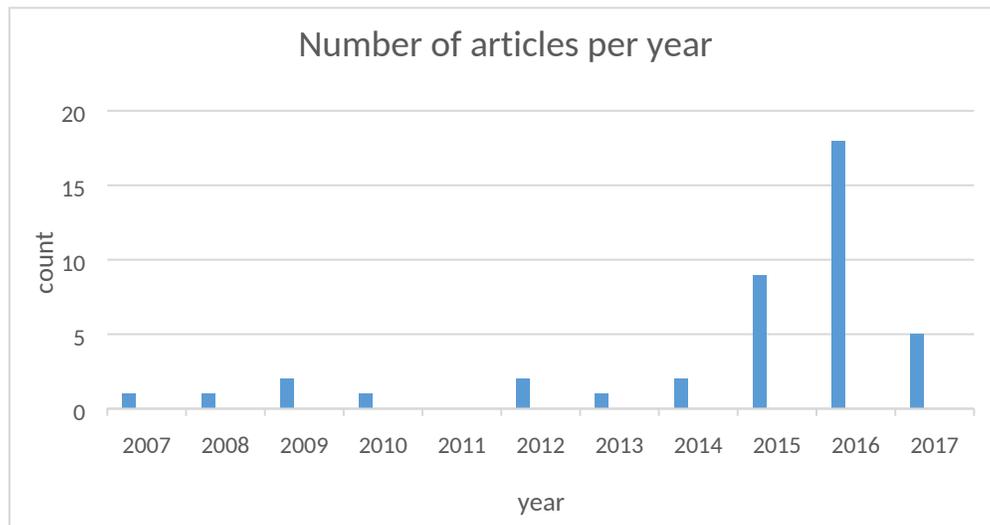

Fig. 2: distribution of obtained article by year

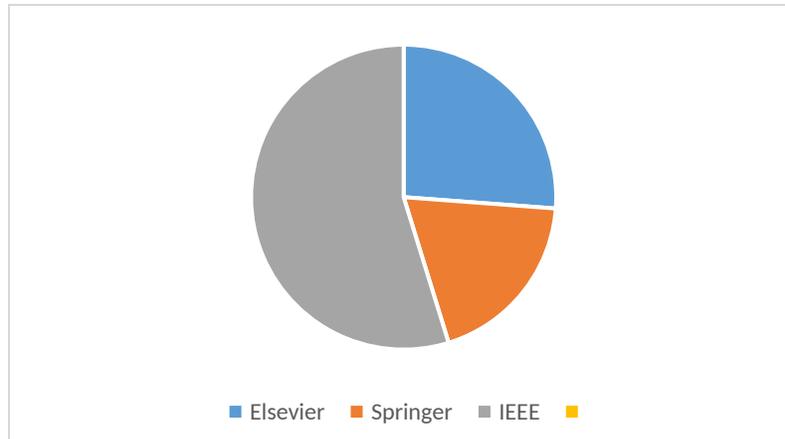

Fig. 3: Number of articles based on publishers

### 4.3. Article Classification

Classifying articles can done based on multiple criteria. In this area, proposed methods about malicious node are different. Some proposed methods of papers only detect malicious nodes. Some papers proposed methods to detect and remove malicious node from network. Above 90 percentage of articles are based on detection of malicious node and articles based on remove malicious node are very low. Actually before removing a malicious node from network, must detect it. Some papers discussed the impact of malicious node to network.

We classified articles based on proposed methods. We first describe detection methods of malicious node that include large number of articles. Little articles cover both subject detection and removal that we describe them. In our article set, there are papers that only explain the impact of malicious node in network and do not proposed any method to detect or remove malicious nodes. We cover this type of papers and describe them.

| # | Year | Name | Criterion | Proposed technique | Based On | Network Type | Attack Type | Parameter | Application Scenario | Simulation Environment |
|---|------|------|-----------|-------------------|----------|--------------|-------------|-----------|----------------------|------------------------|
| 1 | Elsevier,2016 | OMBMND (1) | FNR(False Negative Rate), FPR (False Positive Rate) | | Outlier mining-based | Hybrid P2P | Collusion Attacks and Sybil Attacks | FNR, FPR, top-k (size of LocalFP) | File Sharing Network | |
| 2 | Elsevier,2015 | OODA (3) | reducing the Malicious node detection error | cognitive network management | Observe, Orient, Decide and Act | Heterogeneous Networks | | (DNN-CT), (MFNN) | continuous time dynamic neural network, multi-layer feed forward neural network | |
| 3 | Elsevier,2015 | DSDV (2) | Packet Delivery Ratio (PDR), Delay, Throughput, | I-Watchdog Protocol | DSDV Routing Protocol | Mobile Ad hoc Network (MANET) | | PDR | | Network Simulator (NS-2) Ubuntu Platform |
| 4 | Elsevier,2014 | GTA (17) | Attack Success Rate, attack gain, channel unreliability | Game theoretic (model the malicious node detection game) | Markov Perfect Bayes–Nash Equilibrium | Wireless networks | PBE Attack | | | |
| 5 | Elsevier, 2014 | TEAM (18) | excluding misbehaving nodes | Trust-based Exclusion Access-control Mechanism | based on a trust model | MANET | | Exclusion percentage, time | | Network Simulator 3 (NS-3) |
| 6 | Elsevier, 2013 | RETENTION (5) | reduce the amount of false-negatives | comparison and a performance evaluation | based on trust models | ad hoc grid environments | | Total of punishments | | |
| 7 | Elsevier, 2012 | SPRT (23) | Average number of time slots, False positive rates | Sequential Probability Ratio Test | | static sensor networks | Packet flooding | False positive rates | | |
| 8 | Elsevier, 2008 | IOMN (14) | Quality of Throughput | Calculating trust level of neighbors | Guard node based scheme | Ad hoc On-Demand Distance Vector | impersonation attack, colluding nodes attack, black | trust value of node | 400 _ 400 m cluster of 25 nodes | OPNET |

| | | | | | | hole attack | | | | |
|---|---|---|---|---|---|---|---|---|---|---|
| 9 | Elsevier, 2007 | SecCBSN (7) | Storage overhead, Computation overhead | adaptive security modules (SecCBSN) | dynamic authentication scheme | cluster-based sensor networks | | | | |
| 10 | Springer,2016 | EMATP (19) | packet delivery ratio, average delay and communication overhead, detection accuracy | Enhanced multi-attribute trust | Trust management | WSNs | | | | |
| 11 | Springer,2016 | MATM (19) | End to end delay, Packet delivery ratio, Percentage of malicious nodes detected | Fuzzy logic | based multi-attribute trust model | WSNs | | | | Network Simulator-2 (NS-2) |
| 12 | Springer,2016 | IDS (15) | overcome all the previously specified limitations like ambiguous collisions, receiver collisions, limited transmission power | Intrusion Detection System (IDS) | acknowledgement based intrusion detection system | MANET | | | | |
| 13 | Springer,2016 | MPDP (24) | delivery ratio | a practical approach for cluster head election, | | MANET | | Maximum speed, Data payload, … | | |
| 14 | Springer,2012 | EEBDMN (8) | packet throughput, percentage of byte overheads, energy consumption, and the accuracy of detection | identify the malicious node on forward routing path | energy based scheme for detecting the malicious node | Mobile wireless sensor networks (MWSN) | | | | simulated in glomosim |
| 15 | Springer,2010 | D-CENDA (25) | detection success rate, false positive, false negative | Dynamic Camouflage Event-Based Malicious Node Detection Architecture | Dynamic Camouflage Event-Based | Sensor networks | | | | |

| # | | | | | | | | | | |
|---|---|---|---|---|---|---|---|---|---|---|
| 16 | Springer,2009 | SRBA (26) | throughput | simulate the relevance based approach using NS-2 in a real scenario and consider the impact of malicious node | relevance based approach | VANETs | | | | NS-2 |
| 17 | Springer,2009 | PKIBMRAS (10) | Time | mesh router authentication scheme | Public Key Infrastructure (PKI) Based | Wireless Mesh Network | | | | simulation in C++ |
| 18 | IEEE, 2017 | TESDA (27) | Overhead, Energy Consumption | some methods for secured data transmission | | Wireless sensor networks | Physical Attacks, Interface attacks, Software-Level Attacks | | | |
| 19 | IEEE, 2017 | SIBAD (28) | Incentive | | Secure Incentive Based Advertisement Distribution | VANET | | | | |
| 20 | IEEE, 2017 | CEA (29) | Trust | proposing an algorithm to identify the nodes with highest level of trust within the cluster | based on the calculated trust level | MANET (clustered network) | | | | |
| 21 | IEEE, 2017 | CoCoWa (30) | detection of selfish node | collaborative contact based watchdog | Based watchdog | MANET, DTN, clustered network | | | | |
| 22 | IEEE, 2017 | CRCMD&R (31) | Reputation and legitimacy tables | cooperative malicious node Detection and Removal | Cluster and Reputation based | MANET | worm hole attack | Legitimacy value table and Reputation level table | | |
| 23 | IEEE, 2016 | ECBDS (32) | overhead and end to end delay, Packet Delivery Ratio | design a Dynamic Source routing (DSR) | based energy efficient malicious node detection scheme Extended Cooperative Bait Detection Scheme (ECBDS) | | Black hole and gray hole | PDR, RO | | Network Simulator (NS-2) |
| 24 | IEEE, 2016 | AODRP (9) | Detection of malicious node | detect malicious node in VANET using | | VANET | Sybil Attack | | | NS2 simulator |
| | | | | AODV | | | | | | |
| 25 | IEEE, 2016 | DMNIN (33) | packet delivery ratio, Packet received and lost | malicious node detection based on packet delivery ratio | based on packet delivery ratio | WSN | | PDR | | |
| 26 | IEEE, 2016 | MNDUDST (34) | Packet Delivery Ratio, Throughput, Routing-overhead | Malicious Node Detection Using Dempster Shafer Theory | used SOM classifier, ANN based technique | VANET | DoS | | | |
| 27 | IEEE, 2016 | REABSCSA (35) | Final Target Position Estimation | a secure compressive sensing (CS) based localization approach to cope with the attacked measurements from the malicious node | based Secure CS Approach | WSN | malicious node attack | | | |
| 28 | IEEE, 2016 | BAN-Trust (12) | Precision, Recall | attack-resilient malicious node detection scheme (BAN-Trust) | Based on trustworthiness | Body area networks (BAN) | malicious node attack | | | |
| 29 | IEEE, 2016 | CAMND (36) | Malicious Node Detection accuracy | Propose the cooperative approach for malicious node detection | | MANET | | | | |
| 30 | IEEE, 2016 | ICONFIDANT (4) | throughput, packet delivery ratio (PDR) | Improved CONFIDANT technique | uses DSR routing protocol | Ad-hoc network | | | | |
| 31 | IEEE, 2016 | RBMND (13) | Malicious node detection accuracy, packet dropping rate | Malicious path and packet dropping detection using Merkle trees | REPUTATION BASED | Opportunistic Networks | Packet dropping attacks | | | Opportunistic Network Environment simulator (ONE) |
| 32 | IEEE, 2016 | ESMD (37) | Energy consumption, transmission overhead | Energy efficient and secure data aggregation model | Based on data aggregation | WSN | | | | NS2 |
| 33 | IEEE, 2016 | MNDIDAT (38) | Security, computation | malicious node detection protocol using | Hash-based Authentication, ID-based | ad hoc as well as internet networks | | | | AVISPA |

| | | | | authentication technique | authentication | | | | | |
|---|---|---|---|---|---|---|---|---|---|---|
| 34 | IEEE, 2016 | DOMN&DORS (39) | Packet drop, Throughput, Packet delivery ratio, Packet receive | presents a routing strategy to prevent from attack and identify the malicious node | | VANET | | | | QualNet5.0 |
| 35 | IEEE, 2016 | TKQP&MND (40) | Number of identified malicious nodes, Rate of the misidentification | Top-k Query Processing and Malicious Node Identification | Based on Node Grouping | MANET | new attack model, DRA | | | QualNet5.2 |
| 36 | IEEE, 2015 | LFRS&MN (41) | Broadcast, Throughput, Location, Energy | Revealing Selfish and Malicious Node | | MANET | | | | |
| 37 | IEEE, 2015 | MNTB (42) | Nodes detection accuracy, False negative rate | novel detection and trace back mechanism | Based Merkle tree hashing technique | Opportunistic Networks | packets faking attack | | | |
| 38 | IEEE, 2015 | ZIDS (43) | Accuracy in Detection Rate, Processing Time | uses potential of game theory in order to extracts the uncertain strategies of the malicious node | Zonal-based Intrusion Detection System | MANET | | | | |
| 39 | IEEE, 2015 | DAOMN (44) | route discovery time, throughput | implemented the improved AODV which detects and avoid malicious nodes | Based on AODV | MANET | Resource Consumption Attack, Rushing Attack, Black Hole Attack, Gray hole attack | | | OPNET |
| 40 | IEEE, 2015 | EEDOMN (45) | Network latency | give a secure load balanced node clustering Reliable Node Disjoint Multipath Routing | using trust values of nodes | WSN | | | | Network Simulator NS2 |
| 41 | IEEE, 2015 | EAACK (46) | Detection | proposing an improved | making use of Elliptic curve | MANET | | | | |

| | | | | EAACK scheme | cryptography (ECC) | | | | | |
|---|---|---|---|---|---|---|---|---|---|---|
| 42 | IEEE, 2015 | MNDHCT (47) | Nodes detection accuracy | present a new malicious nodes detection technique against packet faking attack | hash chain based defense techniques | OppNets | Packet dropping attack, packet faking attack | | | |
| 43 | IEEE,2015 | WLAN-AP (48) | detection of malicious node | propose a new protocol for mitigating a bad effects of the malicious node in 802.11 | discard received data from malicious node | 802.11 | ACK-NAV effect | throughput, fairness index, | | QualNet |
| 44 | IEEE, 2015 | MITRM (11) | less time complexity and better accuracy in detection of malicious node | ITRM algorithm where raters are grouped and given a chance before detecting a malicious node | modified iterative trust reputation mechanism | DTN | bad mouthing, ballot stuffing | inconsistency value | | |
| 45 | IEEE, 2014 | top-k QP (49) | identification of malicious node, high accuracy of query result | uses received message information to narrows down the malicious node candidates | top-k query processing | MANET | data replacement attack | accuracy of query result, traffic, identification rate of malicious node | | |
| 46 | IEEE, 2014 | IMTVCM (50) | detect malicious node | maintain a table about node behavior that calculate node trust value | using trust value | MANET | packet dropping | number of packets, transmission range | | NS-2 |
| 47 | IEEE, 2014 | MEC (51) | malicious node detection accuracy | based on maintaining weight of trust value and decrease or increase this value | using minimal event cycle computation | WSN | | number of event cycle, probability of malicious node detection, | | |
| 48 | IEEE,2014 | MAA (52) | detection of malicious node | mobile autonomous agent | agent-based multicast routing scheme | MANET | | packet delivery ratio, power consumption, bandwidth efficiency, | | NS2.34 |
| 49 | IEEE, 2013 | EU (53) | detection | integrating | | WSN | | detection | | |

| | | | | accuracy | wireless sensor network with cloud computing | | | | accuracy, resiliency | | |
|---|---|---|---|---|---|---|---|---|---|---|---|
| 50 | IEEE, 2013 | MND (54) | detection of malicious node | the effect of shadowing and fading on the power signal | base on the received signal strength indicator | WSN | | Sybil, wormhole, dos, malicious data injection, black-hole | neighbor density, transmission power, detection rate, malicious verdict | | MATLAB |
| 51 | IEEE, 2013 | UNS (55) | Identify malicious node | Propose am omniscient algorithm that is capable of tolerating any bias introduced by the adversary in the input steam | Analyze the stationary behavior of this algorithm through a markov chain analysis | | | | Frequency, | | |
| 52 | IEEE, 2013 | CLA (56) | Detection of malicious node | Propose a trust formulation that considers observation, uncertainty, experience as trust value | Trust based | MANET | | Passive attack, active attack | Packet delivery ratio, routing overhead, throughput, end to end delay | | |
| 53 | IEEE, | EPF (57) | Detection of malicious node | Constructs energy field an use an energy prediction model | Based on energy potential field | WSN | | Energy consumption attack | Number of survival node, | | |
| 54 | IEEE, 2012 | AoC (58) | Detection of malicious node and preventive measures | Monitor the transaction among the nodes and detect malicious node | Adoption of cognitive technique | Wireless network | | | Intrusion detection probability | | |
| 55 | IEEE, 2012 | DMDM (59) | Detect the malicious nodes | Use of k-means clustering algorithm and j48 algorithm. | Based on data mining | WSN | | | Dropped packets | | NS2 |
| 56 | IEEE, 2012 | MCMD (60) | Detection of malicious node | Classify nodes to known and unknown types | Based on multivariable classification | WSN | | | Successful sending rate, false detection rate | | |

| | | | | | | | | | | |
|---|---|---|---|---|---|---|---|---|---|---|
| 57 | IEEE, 2012 | PPN (61) | Detection and removal | Each node in the network has an identity that can detect malicious node | Based on node identity and AODV Protocol | MANET | Black hole and gray hole attack | Throughput | | |
| 58 | IEEE, 2011 | RNS (62) | Detection | Propose a robust sensing node selection algorithm which evaluate trustworthiness of nodes | Based on trustworthiness of nodes | | | Probability of miss detection | | |
| 59 | IEEE, 2011 | VMAT(63) | Malicious sensor revocation | Secure aggregation protocol with malicious sensor revocation capability | Using symmetric key cryptography | WSN | | Number of sensor mis-revoked, error | | |
| 60 | IEEE, 2011 | CBDS (64) | Prevent malicious node attacks | Cooperative bait detection schema | | MANET | Black hole, gray hole | Packet delivery ratio, overhead ratio | | |
| 61 | IEEE, 2010 | PLD (65) | Malicious node detection | Received signal strength in physical layer is used for detection | Via Physical layer data trust metrics | Sensor networks | | | | |
| 62 | IEEE, 2009 | CENDA (66) | Detection rate | Camouflage event based malicious node detection architecture | Uses camouflage events generated by mobile-nodes to detect malicious nodes | | Sybil attack, wormhole attack | Detection success rate, false positive, false negative | | |
| 63 | IEEE, 2009 | BAMN (67) | Throughput | Analyze the effect of black hole attack | | MANET | Black hole attack, | Data lose | | NS2 |
| 64 | IEEE, 2009 | CMND (68) | Discover malicious nodes | A combined strategy that uses Self-destruction node technique | Based on autoregression technique | WSN | | Error | | |
| 65 | IEEE, 2009 | EMND (69) | | To detect when a node physically measures wrong values | | WSN | | | | |
| **66** | IEEE, 2008 | NRM (70) | Detection rate of malicious node | The beta distribution is used to describe the reputation distribution | Reputation based model | WSN | External attacks, internal attacks | Detection rate | | |
| **67** | IEEE,2008 | ML (71) | History of transaction data | Using reputation system and historical data to | Machine Learning based | | | Node classification error, | | |
| **68** | IEEE, 2007 | ARM (72) | Time series of measured data | Autoregressive prediction for malicious nodes detection | Based on past/present value of each node | WSN | Spoofing, selective forwarding, sinkhole attack, wormhole attack | Time series | | |
| **69** | IEEE, 2007 | NCP (73) | Statement a node about another node | A good node make statement about another node | Based on court work | | | | | |
| **70** | IEEE, 2004 | MSDSS (74) | Signal strength | The idea is to compare the signal strength of a reception with it expected value, calculated using geographical information | Signal strength of message transmission | WSN | HELLO flood attack, wormhole attack | network density, message check probability, transmission power, maximum ratio difference | | |

## 5. Article Grouping

In this section, we categorized studied papers in some groups and describe each paper in general description of their performance. The grouping of papers is as below and some paper that does not fall any category, described separate.

### 1. Routing Protocol based Algorithms

In this title, we describe papers that benefit from routing protocols. These methods presents techniques with minor change in routing protocols like DSR, DSDV and so on to prevent malicious nodes from participate in routing.

(32) Designed a malicious node identification based on dynamic source routing. This paper decreased routing overhead and end-to-end delay. This paper only detects malicious nodes. These malicious nodes identified used reverse tracking. This work presents a malicious node detection mechanism named Extended Cooperative Bait Detection Scheme (ECBDS) that combine with proactive and reactive defense architecture and create high efficiency.

(9) Detects malicious nodes using AODV (Ad Hoc on Demand Distance Vector) routing protocol. This paper used NS2 simulator. This work uses AODV protocol simulation to identify malicious nodes. This paper used 1000 * 1000 area that some nodes considered as RSU and two node considered as malicious nodes. In this simulation, when a node comes to network, get an id but malicious nodes try to change their id that used this parameter to identify malicious nodes.

(4) Use improved CONFIDANT (Cooperation of nodes fairness in Dynamic Ad Hoc Network) to detect malicious nodes. This improvement caused alongside detecting malicious nodes, prevented from blackhole nodes, false reporting problem solved, network collision problem solved and provide QoS. This way uses DSR Routing protocol and cause improve throughput and PDR. This paper split up nodes to cooperative and noncooperative. A module named monitor module, monitors packet sending.

(39) Presented a routing strategy to prevent malicious node attack and detect malicious nodes. This strategy implemented in QuaNet 5.0. This work proposed in VANET. This routing protocol act using double acknowledgment between sender and receiver. In this method, when S sends a packet to N1, tell to S that who received packet and when N1 for example sends packet to N2, again tell to S that who received packet, in this method if a node could not receive a packet instantly selects alternative node.

(44) Paper improved AODV routing protocol that detected malicious nodes and avoided them. The basis of this work is that first count all packets send by a node, which act as a router and count all dropped packets. In routing process, the value of dropped packets evaluated and based on decided the node be in route or not.

### 2. Ack based Algoeithms

(46) Uses Elliptic curve cryptography for improved security with reduced key size and with less computation. In this study author proposed an improvement in end-to-end acknowledgment scheme called TWOACK. The TWOACK scheme detects malicious links not nodes. This method uses acknowledgment technique. This article has two modes that are same as EAACK and uses additional mode called special mode. First mode is ACK mode that is like end-to-end acknowledgement scheme. After a predefined time, S-ACK mode is activating that is similar to TWOACK. In this mode, group of three consecutive nodes work cooperatively to find misbehaving node. In special mode, source sends special packet directly to next node of the node, which has sent an NACK and waits for special acknowledgement packet from it.

(48) Proposed a new protocol for mitigating the bad effect of the malicious nodes in 802.11 based wireless AP on public spaces. WLAN AP checks the value of duration field when the reception of the packet from client terminal. at this time WLAN AP checks more flag for check whether the fragmentation transmission is applied to the corresponding transmission, if the flag is true and the duration value is not 0, the WLAN AP decide the ACK sender is malicious node.

### 3. Algorithms based on Authentication, Encryption, Hashing

(7) Detects malicious nodes. This study presents an adaptive security module for improvement secure communication to authentication new nodes, create a secure link, and broadcast it between neighbors. This work prevents entering outer malicious nodes. Also, make sure from internal nodes using trust evaluation and black and white lists.

(38) Detects malicious nodes using authentication technique. This paper simulated using AVISPA software. This work introduced an ID-Based lightweight encryption protocol using authentication technique. Proposed technique is based on one-way hash function and not need more high computation. This method used a temporary server for authentication.

(42) Presented a method in Oppnets for detecting malicious nodes and evaluate a packet dropping attack. This paper also presented a tracing mechanism. The method used in this paper is for merkle tree hashing. When packets received to legal node, the node could create merkle root hash value and compare it with received. The attack that this paper analyzed that is it a malicious node drop packet and inject a fake packet.

(47) Uses hash chain based defense technique that have two phases. The first phase is to detect the attack and the second phase is to find the malicious nodes. Sender node calculates the chain values for the packets based on packet contents and then automatically include them with each packet. The receiver node recalculates hash values and compares them with received hash values. If calculated values were same as received values, chain doesn't break and no malicious node detected.

(63) propose a novel secure aggregation protocol called VMAT (verifiable minimum with audit trail), using only symmetric key cryptography. To avoid the overhead of using public key cryptography on sensors, VMAT will only use symmetric key cryptography. In VMAT, each sensor shares a unique symmetric key with the base station. If a sensor is destroyed or radio-jammed by the adversary, considered it as being malicious. First a tree formation phase conducted and aggregation tree be built.

(10) Used public key infrastructure to protect network from malicious nodes by authentication method. This method increase end-to-end delay and can detect malicious nodes and delete it from network. This paper produced a router mesh authentication method that solve problems.

4. **Message Communicating based Schemes**

(60) Basic idea is to store the basic message about communications between nodes, and send them to the base node where these messages are combined to node feature vectors. Once the base node gets all nodes feature vector, it uses the type known nodes as a training set, generate a classifier with the multivariable classification algorithm, and labels the type-unknown nodes.

(74) Provides a solution to identify malicious nodes through detection of malicious message transmissions in a network. A message transmission is considered suspicious if its signal strength is incompatible with its originator's geographical position. A transmission is malicious if the geographical position included in the corresponding message is made up or was transmitted with a power that differs from the one agreed upon by all the other nodes in the system. After checking the suspiciousness of a received message, the node updates its table accordingly: if the message is suspicious, it increase the message originator's suspicious count by one.

5. **Trust, Voting and Reputation Methods**

(18) Used local and global information of nodes and a voting system to exclude malicious nodes from network. This paper actually mixed mechanism of trust based and voting based to delete malicious nodes from the network. This mechanism divides access control responsible in two section: local and global. The responsible of local section is consideration of neighbors to send to the global section. As a result global section analyses received information and used of a voting system, punishment the suspicious node.

(5) Detects malicious nodes using trust model and punish them. This paper detects all 100% of malicious nodes and punish them without create false positive. This paper divides confronting the attack techniques in two groups, preventive mechanisms and reactive mechanisms that detecting malicious nodes do in reactive mechanisms section and this paper focuses to this section. The most popular reactive system is IDS. The punishment system is in the form of local and distributed that in local mode done by user and in distributed mode done by other users.

(14) Identifies malicious nodes. Guard nodes do this work. This paper also is based on trust level of neighbor nodes. This technique remove malicious nodes after identify them. In this scheme, every node calculate trust level of neighbor nodes to select route. If trust level of a neighbor were lower than a threshold value, that node identified as a malicious node and not considered in route selection.

(29) Produce an algorithm that selects a node with highest trust in cluster as a cluster head in order to not selected malicious and selfish nodes as a cluster head. In this work, the nodes with lowest trust placed in malicious nodes list. The trust of users specified based on their activity and inserted in a trust table. The trust level specified based on residual energy and the role of node in packet forwarding. The level of node interaction with neighbors considered as a trust increase factor.

(31) Used cluster and reputation to produce a technique for detect and remove malicious nodes. This work used a reputation table for each node. CRCMD&R scheme is a protocol like AODV based on demand that avoid malicious nodes in creating route. Each Cluster head has a neighbor table for each neighbor and a table for trust level that record information about all nodes in it. In route selection, routes that trust level is not lower that a threshold. In addition, this work uses a table named legitimacy table that shows history of successful routes and selection of routes is based on.

(12) Paper used an attack-resilient malicious nodes detection in body area networks (BAN) and detected malicious nodes. The algorithm used in this paper is BAN-Trust. This is done used viewing behavior of nodes. Reliability of each node calculated based on historical node behavior. This paper has two steps, first analyzed data and then send to trust management to detect malicious nodes.

(13) Presents a technique to detect malicious nodes, malicious routes and packet dropping. Merkle tree hashing used to detecting packet dropping that create trust for each node and using it malicious node identified. This paper is reputation based. When a reputation of a node is lower than a threshold, considered as a malicious node.

(45) Try to decrease the effect of malicious nodes by balancing values of trust and residual energy and improve network lifetime. This work present a balanced node clustering using trust values, also by creating backup cluster head try to improve reliability in the network. First, cluster head calculates the average residual energy of nodes by broadcasting head message. In this paper, trust value calculated using received acknowledgment packets. The node with highest trust value and highest residual energy selected as cluster head and this cause the nodes with low energy and low trust value don't participate in routing.

(50) Proposed work of this article identifies malicious node using trust value in cluster based MANET is used to detect the malicious node in cluster. In this method, the nodes direct and indirect trust values are used to detect the malicious node in the network. Nearby node gives the direct trust value about a node and the cluster head also ask the other members of the cluster about the misbehaving node. The other members have trust value about that node means; it sends the indirect trust value to the cluster head. By this way the malicious node are identified in the cluster in efficient manner. This work has two phases, cluster formation and trust value calculation. The direct trust value and the indirect trust value are collected from cluster members by the cluster head.

(51) Proposed a method that detect malicious node by computing the average number of event cycles. Each sensor node maintains a list of its neighbors and their trust values. The weight W represents the own trust value ranging from 0 and 1. Malicious node in an event region reporting 0 intentionally to lead an incorrect decision. By the formula proposed, at each event cycle, a malicious node losses its weight by a with probability P or gain its weight by b with probability 1-P. The node will be determined to be malicious at the time of weight, initiated to 1, reaches 0.

(56) Proposed a trust based technique that uses multiple parameters with different weights. This approach considers observation, uncertainty, experience, recommendation and correctness of recommendation while calculating trust value for a node.

(62) Proposed a robust sensing node selection algorithm which evaluates the level of trustworthiness via a weighted consistency check with reports of trusted nodes.

(19) Used an intrusion detection system based on trust that uses multi-attribute trust metrics to improve detection accuracy. This paper only detects malicious nodes. This work uses a trust calculation algorithm that include monitor neighbor nodes and trust calculation using Message Success Rate (MSR), Elapsed Time at Node (ETN), Correctness (CS) and Fairness (FS). This paper considered cluster architecture and has cluster heads, Sensor nodes and base station.

For prevent from attacks, two method include prevention based and detection based can be used. Prevention based methods include (encryption, authentication protocols and access controls) that rely on centralized infrastructure and contract in first level, and in second level are detection based approaches that there are an Intrusion Detection System (IDS) based on trust. IDS approach highly identify behavior of nodes. IDS works in two way: anomaly detection and misuse detection, which this paper is based on anomaly.

(19) Used trust model based on fuzzy logistic. Used elapsed time at node, correctness and fairness as trust metrics. This work uses fuzzy calculation theory to calculate final trust value. This paper in security area, only detects malicious nodes.

(71) Propose a machine learning based Reputation system that automates the process of devising the reputation system model and defends against many patterns of attacks. Given the history of transaction data, the goal would be to construct machine-learning classifiers that can utilize this data to make predictions. This article envision the problem of reputation system as a time series prediction problem.

### 6. Throughput and Packet Delivery based

(33) Using Packet Delivery Ratio (PDR) criterion detects malicious nodes. The packets that created in source must equal the packets received in destination, so not being equal meaning malicious nodes are in the route.

(26) Only considered the effect of throughput in VANET and did not do anything about detecting or removing malicious nodes. This paper considered relationship based approach where in sharing information rely on intermediate nodes.

### 7. Based on Spatial and Temporal Information

(66) Exploit the spatial and temporal information of camouflage event while analyzing the packets to identify malicious activity. A camouflage event is a reputable event generated in response to a basestation request. A camouflage event generator is a mobile-node, which can be mounted on robot or an unmanned aerial vehicle. This model provide a lightweight address encoding scheme wherein each node encodes the relative address of the node from which it received the packet.

(25) Produced an architecture that detects malicious nodes and in addition identify type of attack. This method used spatial and temporal information for detecting malicious activity. This work is done by using base station information. A node as a camouflage node can gather information about other nodes. Gathered information by camouflaged node send to base station.

(35) Used a safe compression sensitivity (CS) to locate malicious node attack with analyzing remaining error. This work used CS technique and spatial information to confront malicious node attack. First using CS algorithm estimated target position and then calculates residual error and find out calculations of suspicious attacks.

### 8. Based on Node Behavior

(1) Detects malicious nodes with outlier mining-based technique in P2P networks. This work is done with behavior patterns of nodes. Simulation results demonstrated that this technique detects malicious nodes with lower false positive rate and false negative rate. This paper is used super node and ordinary node information that are in hybrid P2P networks. The super nodes are responsible for transfer and maintaining interaction data that used for detecting malicious nodes. This paper also benefit from feedback trust model for detecting malicious nodes. For each subnet that consist of multiple ordinary node and managed with a super node, extracted alternate behavior of pattern and with this information detects malicious nodes.

(2) Detects malicious nodes in the network using improvement of watchdog protocol. This paper focuses on DSDV routing protocol. In this paper, presented I-watchdog procedure exhibit effective detection of presence of malicious nodes in mobile ad hoc networks and find out the reason of packet loss. Besides that, analyses improving performance of mobile ad hoc networks in presence of DSDV with I-watchdog protocol from the point of packet drop ration (PDR) through end-to-end delay.

(23) Try to detect malicious nodes and deprive them from continue to communication. This paper apply a sequential hypothesis testing to discovery the nodes that are silent for a long time and block their communication. By detect and blocking in local mode, energy consumption overhead placed in lowest.

(30) Using classify network to clusters so that each cluster have a cluster head and identify behavior of nodes to not behave selfish of malicious. This work done by watchdog protocol. This paper only detects selfish nodes. Supervised buffer level of all nodes to identified packet loss level and detects malicious node in this way.

(74) Proposed a decentralized malicious node detection technique based on the received signal strength indicator. The effect of shadowing and fading on the power signal in wireless channels cannot be discarded and used in this technique. When a victim node receives a packet, it decrypts the coordinates field using the shared key and calculates the accepted RSSI range of values for the computed distance. the detection module compares the actual received power value with the calculated accepted range. If the received value falls within the range, the packet is accepted. otherwise, the node creates a malicious-verdict and transmits it to a number of neighboring nodes to perform the detection process.

The proposed method of (59) makes use of k-means clustering algorithm and J48 algorithm. The basic motivation to proposed approach is to detect the malicious node in sensor node environment. The basic idea is that analyze the behavior of each of the nodes in the network, if a node dropped all the information, which implies that a node has been compromised or malicious. The classification is done by J48 algorithm for the formulation of the decision tree.

In (70), the beta distribution is used to describe the reputation distribution. According to the characteristics of various internal attacks, the anomaly behavior detection module establishes a series of anomaly-detection rules and behavioral parameter statistical rules for judging whether abnormal events happen on nodes. In this paper, reputation evaluation module is responsible for the integration and calculation of node's behavioral parameter, which are the output of anomaly detection module.

(72) Compares at each moment the sensor's output with its estimated value computed by an autoregressive predictor. A malicious sensor node that will try to enter false information into the sensor network will be identified by comparing its output value x(t) with the value x'(t) predicted using past/present values provided by the same sensor. First describes a threshold that will be used to determine if a sensor acts normal or abnormal.

## 9. Based on Grouping and Clustering

(36) Wants to detect malicious nodes in MANET. This work is done by selecting appropriate cluster head to increase performance. Cluster head could delete RREQ in reason low energy and send a RELEASE message to all nodes, when a node received a RELEASE message, remove being cluster head request from queue and this help to malicious nodes detection.

(37) Identify malicious nodes to decrease energy consumption by nodes. In sensor networks that gathered information send to collector node, the collector node must not be malicious. Nodes are divided to leaf nodes, intermediate nodes and sink nodes. Gathered data send to parent node and parent node again calculate data gathering, if calculated values are not equal to received values, the node considered as malicious node.

(40) Produced a method for identifying malicious nodes in MANET using node grouping. After detecting malicious node, this node declared to other nodes using message exchange. For comfort in this work, nodes are grouped. QuaNet 5.2 used for simulation. Top-k query protocol used in this paper and there is a node that do query issuing. In this network, malicious node try to disrupt in query issuing. This paper presented a new attack name Data Replacement Attack (DRA) that replace malicious data. This paper also presented a new attack named False Notification Attack (FNA) that a malicious node introduce a normal node as a malicious node.

(41) produced a method for malicious node detection, cluster head detection, and discussed about overcome to energy consumption. This paper presented its method in MANET. This paper used

Aloha protocol for media access control. This paper used information send by cluster head to detect malicious nodes.

(27) Detects malicious nodes and isolate them. This technique sends received information from nodes to base station to be processed.

### 10. Based on Artificial Neural Network

(34) Identify malicious nodes using artificial neural networks. This work identify malicious nodes in VANET. The attack that considered for this network is DOS attack. Targets of this paper are development a secure communication network and development a combined technique to packet filtering. This paper uses self-Organizing Map (SOM) that have a neural network which learn input data features from neighbors. SOM grouping nodes and analyze behavior of nodes to detect malicious nodes.

(58) Proposed a technique that monitor the transactions among the nodes in the network and detects the malicious nodes and take preventive measures. This paper used single sensing with cognition and training of network using artificial neural network to achieve high detection rate. To achieve cognition in wireless networks, this paper proposed to use back propagation algorithm for learning and to observe the behavior of nodes. Numerous pre-defined data transactions are carried out and these transactions need to be monitored and stored in observed node behavior section of cognitive framework to detect malicious nodes.

### 11. Energy Based

(57) Propose a detection method of malicious nodes of wireless sensor network based on energy field. First, constructing wireless sensor network based on energy field through the energy potential field of the routing protocol. This article uses an energy prediction model that can balance the network load and detect the malicious nodes in network correctly, so that the network can avoid the threat from malicious nodes.

(8) Works based on detecting malicious nodes. Proposed method of this paper evaluated based on packet throughput, percentage of byte overhead, energy consumption and accuracy. This paper focused energy based schemes for detecting malicious nodes.

### 12. Based on Game Theory

(17) Used game theory for interaction between normal and malicious nodes to detect and coexistence. This article after detecting malicious nodes, create a post-detection game between normal and malicious nodes to happen coexistence. This paper uses a special node to monitor behavior of nodes to detection of behavior malicious nodes.

(43) Used game theory potential to discover unknown strategy of malicious nodes. This paper is considered a zonal-based intrusion detection in MANET. This method present a high detect accuracy. The proposed model uses mobile node detection based on behavior. This method used multi stage game theory. Considers every mobile node as an actor in game and consider sets of actions for every actor that combination of every action is as mobile nodes strategy.

### 13. Node ID Based

(52) Proposed an algorithm to find link failures by acquiring information from source and destination using agents and repairing the link. The source and destination and the member nodes are assigned with its unique id so that the nodes can be uniquely identified. Intrusion detection system detects malicious activity. The application of the various rules incorporated in the mobile agents. mobile agents will work as a carrier for id's to move to a particular node and collect information from the node this information is transferred to the base station.

(61) Prime Product Number scheme is proposed to mitigate the adverse effects of misbehaving nodes. The basic idea of the PPN scheme is that, each node in the network has a specific prime number which acts as Node Identity and this identity must not be changed. This scheme is based on AODV and every cluster head node maintain the neighbor table, which is used to keep information about all the nodes. An intermediate node will attempt to create a route that does not go through a node whose replied information is wrong and PPN is not divisible. Therefore, malicious nodes will be gradually avoided by other non-malicious nodes in the network.

In the malicious node removal process respective CH add malicious node to the malicious list and broadcast this to the whole network.

### 14. Other Methods

(3) Detects malicious nodes and malicious behavior with cognitive-based methods (observe, orient, decide and act). Actually, this article presents a network management technique based on cognitive. In this method, the nodes always monitored and in case of seeing suspicious do proper planning. This article considered heterogeneous networks.

(28) Identify malicious nodes and only analyze effect of them to VANET. This paper uses advertisements displayed in roads for driver. The objects participant in this scheme are advertisement content server (ACS), Advertisement Distribution Point (ADP), and vehicles. This scheme shows set of nodes that have malicious behavior and analyze the effect of them to SIBAD. In addition, this scheme uses two steps verification.

(11) Modified ITRM algorithm. In ITRM algorithm, iterative transactions take place between the service providers and raters and the raters give rating to the SPs after every transaction. Existing ITRM algorithm detects honest raters as a malicious node and hence results in more false positives. In this method the raters are divided into three types of clusters based on rating given to service providers such as high priority, middle priority and low priority. The algorithm upgrades the rater from low to high priority cluster and vice versa depending on rating given by him instead of blacklisting rater immediately.

(49) Proposes methods for top-k query processing and malicious node identification against data replacement attack. In the top-k query processing method, in order to maintain accuracy of the query result, nodes reply with data items with the k highest scores, along multiple routes. In this method, first the query-issuing node floods a query over the entire network. After detecting an attack, the query-issuing node identifies the malicious node. The query-issuing node narrows down the candidates for the malicious node, and identifies the malicious node by making respective inquiries.

(53) Focusing on investigating integration wireless sensor networks with cloud computing. the objective of this work is to determine the optimal sensor placement that would maximize the highest ration of the signal/noise by which can control the sensor states.

(55) First propose an omniscient strategy that processes on the fly an unbounded and arbitrarily biased input stream made of node identifiers exchanged within the system, and outputs a stream that preserves uniformity and freshness properties. Proposed am omniscient algorithm called omniscient strategy, that is capable of tolerating any bias introduced by the adversary in the input stream. Then propose a randomized approximation algorithm called knowledge-free strategy that is capable of outgoing an unbiased and non static sample of the population.

(64) Proposed a malicious node detection scheme, named CBDS, which is able to detect and prevent malicious nodes launching black/gray hole attacks and cooperative black hole attacks. It integrates the proactive and reactive defense architecture. By using the address of the adjacent node as the bait destination address, it baits malicious nodes to reply RREP and detects the malicious nodes.

(65) Develops physical layer based trust metrics to detect non-functioning nodes. At the physical layer, wireless network discovery (WND) includes modeling the waveform and modulation type of each transmitter in the non-cooperative network. This paper discusses a physical layer trust model, which is assessed via the following candidate information metrics: consistency: whether or not the information from a sensor agrees with other information in the system. Honesty: this model whether inconsistencies are intentional or not.

(68) Identifies malicious sensors by using an autoregression technique and eliminates them starting a self-destruction node procedure. This paper is based on the detection of malicious sensor nodes using AR predictors and the elimination of their effects by expelling them from the network using a self-destruction node technique. This strategy exploits the temporal redundancy in the sense that provided measurements from each sensor.

(73) Is based on court work. A good node makes a statement on another node. Testifying that it is a good node; the testified node must also be good and accusing that it is a bad node; the accused node must be bad.

(15) Detects malicious nodes in MANET. This work detects malicious nodes and broadcast them. This paper overcome defects of previous works by improving IDS.

**Percentage of Methods**

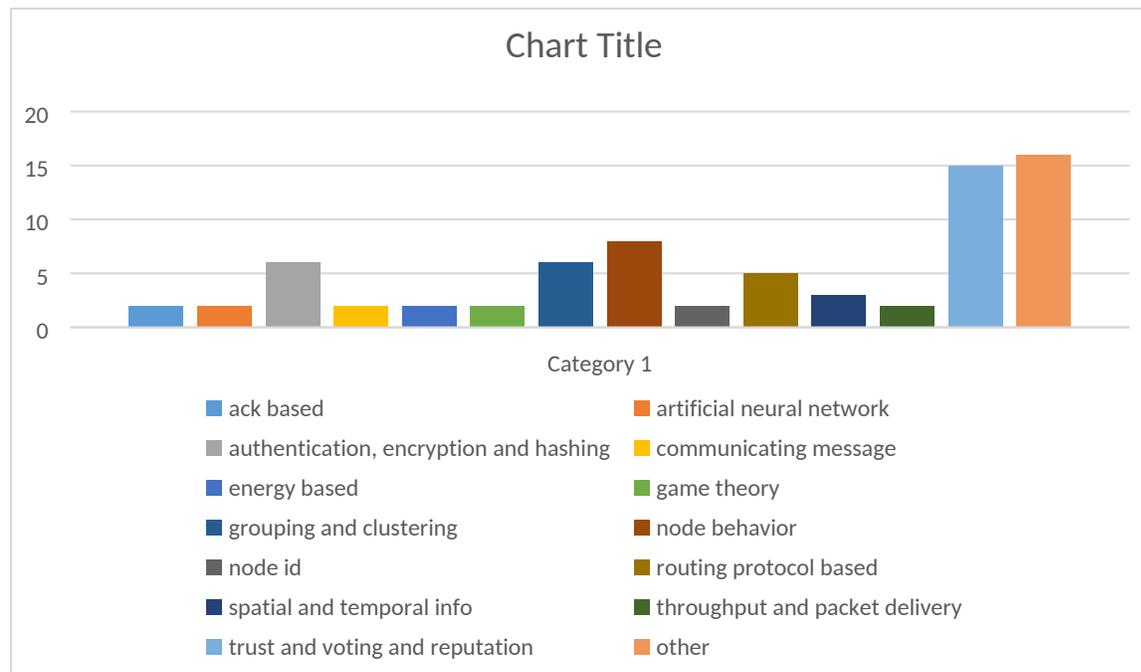

## 6. Conclusion and future works

In this study we read about 70 papers on malicious node detection and extract their methods. We categorized used methods in some groups and compared those methods in table. The techniques based on authentication, encryption and grouping, clustering and node behavior they had most papers and proposed methods in literature. Other techniques like artificial neural network or game theory methods have work place and are not saturated. For future works researchers can benefit from artificial neural network techniques to detect malicious nodes. In game theory methods researchers can use other game models like symmetric asymmetric models, zero-sum game models and other game models that are in game theory.